# Oxygen-driven enhancement of electron correlation in hexagonal iron at Earth's inner core conditions

Bo Gyu Jang[1,2], Yu He[1], Ji Hoon Shim[3,4], Ho-kwang Mao[1], and Duck Young Kim[1*]

[1]Center for High Pressure Science and Technology Advanced Research (HPSTAR), Shanghai 201203, China.

[2]Korea Institute for Advanced Study, Seoul 02455, Republic of Korea

[3]Department of Chemistry, Pohang University of Science and Technology, Pohang 37673, Republic of Korea

[4]Division of Advanced Materials Science, Pohang University of Science and Technology, Pohang 37673, Republic of Korea

*email: duckyoung.kim@hpstar.ac.cn

## Abstract

Earth's inner core consists of mainly iron with a bit of light elements. Understanding of its structure and related physical properties has been elusive for both experiment and theory due to its required extremely high pressure and temperature conditions. Here, using density functional theory plus dynamical mean field theory, we demonstrate that oxygen atoms energetically stabilize hexagonal structured iron at the inner core condition. Electrical resistivity is much enhanced compared with pure hcp-Fe, supporting the conventional thermal convection model. Moreover, our calculated seismic velocity shows a quantitative match with geologically observed Preliminary Reference Earth Model data.

The structure and composition of Earth's core have long been a question of great interest. Although iron is generally believed to be the primary component of the Earth's core, even the phase of iron is still controversial. hcp-Fe is traditionally considered as the stable phase under the inner core (IC) conditions *(1)*. The elastic anisotropy of IC is regarded to be originated from a preferred orientation of hcp-Fe with a non-ideal *c/a* ratio *(2)*. However, it was later shown that the *c/a* ratio of hcp-Fe becomes almost ideal at high temperature, resulting in the vanishing of the anisotropy *(3, 4)*. An alternative scenario based on body-centered cubic (bcc) Fe was proposed *(3, 5, 6)*, in which a diffusion mechanism at high temperatures can make bcc-Fe stable and give anisotropy *(6)*.

The phase of iron at IC conditions is fundamentally important to model Earth's geodynamo mechanism, since the electrical properties of hcp- and bcc-Fe are known to be very different. Electronic correlations effect of transition metals, especially iron, should be considered properly at high temperature conditions *(7-10)* as a previous theoretical study found that the electrical conductivity from the electron-electron (*e-e*) scattering reaches ~35% of that from the electron-phonon (*e-ph*) scattering in case of hcp-Fe at Earth's core conditions *(8)*. The electron correlation effect of bcc-Fe is much larger than that of hcp-Fe and bcc-Fe shows non-Fermi-liquid behavior (or fully incoherent behavior) at IC conditions *(9)*. This implies that the *e-e* scattering part can be even more significant in bcc-Fe case and therefore, the corresponding thermal conductivity should be affected by the structure of iron.

Density and velocity of IC in the Preliminary Reference Earth Model (PREM) *(11)* are smaller than those of pure iron. It is generally accepted that light elements should be involved to explain the density and velocity deficit of IC *(12)*. Although several light elements, such as S, O, Si, C, and H, have been suggested *(13–18)*, it is still a matter of debate because of the lack of direct evidence. Among those elements, oxygen has recently attracted attention as a prime candidate. The discovery of oxygen-rich iron compound, $FeO_2$ under deep lower mantle conditions, suggests that there is more oxygen under deep Earth conditions than we traditionally believed *(19–21)*. Furthermore, the FeO partitioning model implied the potential existence of stable oxygen-enriched layers below the core-mantle boundary *(22-24)*. These recent studies support that oxygen should be properly considered in the Earth's core model.

Although these light elements' effects on velocity deficit and conductivity from *e-ph* scattering have been studied from both the experimental and theoretical sides, there has been no detailed investigation of light elements' effect on the phase of iron and conductivity from *e-e* scattering at IC conditions. For the realistic understanding of the thermal convection of IC, however, the contribution of light elements should be considered properly. The electrical conductivity from *e-e* scattering (electronic correlation effect) can be sensitively affected by not only the phase of iron but also the volatiles in the compounds. Considering that iron oxides are strongly correlated compared with pure iron, the inclusion of oxygen in the IC model can modify significantly previous theoretical estimations on the thermal conductivity.

Here, we investigate the role of a small amount of oxygen on the structural and physical properties of Earth's IC. We find that $Fe_xO$ ($x \geq 3$) can be stabilized under IC conditions using *ab initio* calculations. This series of new iron oxide has a universal hcp-Fe structure motif with intercalated oxygen atom, indicating that a small amount of oxygen can stabilize the hcp phase at IC conditions. Intercalated oxygen atoms can make covalent bonds with Fe atoms and this interaction can explain the anisotropy of the hcp phase. The existence of oxygen also enhances the electronic correlation effect compared to pure hcp-Fe. As a result, the electrical resistivity from the *e-e* scattering part can be much enhanced, indicating that the thermal conductivity can be smaller than the previous estimation on pure iron. The density and velocity deficit of the PREM model are also well described by the existence of oxygen in the hcp-Fe motif.

We first performed structure prediction simulation of the Fe-O system at 300 GPa using the ab initio random structure searching (AIRSS) method based on density functional theory (DFT) calculations (Fig. 1A). We found that $Fe_xO$ ($x \geq 3$) are on the convex hull line, showing energetic stability at IC pressure. From $Fe_3O$ to $Fe_9O$, they share a common structural motif where the oxygen atom is intercalated between Fe atoms with a hcp structure, implying that X-ray diffraction (XRD) patterns of $Fe_xO$ series can be equivalent to that of pure hcp-Fe. $Fe_xO$ possesses *P3m1* space group for $x=2n$ ($x > 1$) and *P$\bar{6}$m2* space group for $x=2n+1$ ($x \geq 1$) (Fig. 1B), respectively. For $x=4n+1$ ($x \geq 1$), strictly speaking, it possesses *Amm2* space group which is equivalent to a unit cell doubling of *P$\bar{6}$m2* structure with a tiny distortion. The energy difference between *P$\bar{6}$m2* and *Amm2* structures is merely ~5 meV/f.u. for $x=5, 7, 9$. It is worth noting that our predicted crystal structures of $x=2, 3$, and 9 cases are in good agreement with previous theoretical studies *(25, 26)*.

Standard DFT is not enough to describe the electron correlation effect of Fe 3*d* orbitals. While one might expect the negligible contribution of the correlation effect with pressure, several studies claimed that its effect is critical to describe the physical properties of iron even at the IC conditions *(7-10)*. For example, previous DFT calculations predicted that a *R3m* structure becomes the ground state of FeO at 300 GPa *(25,26)* but experiments data reported NiAs-type B8 phase at low temperature and CsCl-type B2 phase above 3000 K *(27)*, clearly yielding an inconsistency.

To estimate the correlation effect of Fe 3*d* orbitals on the energetic stability, we performed total energy calculations and revised the convex hull using dynamical mean field theory (DMFT) calculation combined with DFT. Interestingly, DFT+DMFT calculations give an excellent match with experimental data on determining FeO phase, exhibiting the B2 phase becomes the most stable one at 300 GPa (Fig. S1). These results indicate the importance of the electron correlation effect on the physical properties of iron oxides, including the formation energy. Furthermore, we found that hcp-Fe is still energetically favorable than bcc-Fe by ~ 1eV/f. u. (Fig. 1A), which is consistent with DFT calculations and the energy difference is unlikely to be overcome by temperature. Hexagonal $Fe_xO$ series is on the convex hull line from $Fe_3O$ up to hcp-Fe, indicating the stability of $Fe_xO$ ($3 \leq x \leq 9$) under IC conditions (Fig. 1A). Thus, hexagonal-structured iron oxides are preferred in the presence of a small amount of oxygen.

As a representative example of our predicted iron oxide, figure 2A shows the crystal structure of $Fe_9O$. The distance between Fe atoms is aligned along the $c$ direction and it is similar to that of hcp-Fe (black arrows), except for Fe atoms adjacent to the oxygen atom. The distance between Fe atoms above and below the O atom (~2.60 Å) is much shorter than the others, indicating the strong interaction between Fe and O atom. One can find that the O atom is properly screened by the nearest Fe atoms and Fe-Fe distance is fully recovered to ~ 3.40 Å of pure hcp-Fe.

$d$ electrons occupancy of Fe atoms as a function of the distance from O atom is obtained using DFT+DMFT calculations (Fig. 2B). The numbers on Fe atoms in Fig. 2A are assigned in order of the distance from O atom, which correspond to the $y$ ticks in Fig. 2B. The $d$ occupancy of $Fe_xO$ is remained to be smaller than that of hcp-Fe and it monotonically increases as $x$ increases, and it eventually converges to that of hcp-Fe (the vertical dashed line). The occupancy of Fe1 atom is smaller than that of others due to the charge transferring from Fe1 atom to O atom. From the second nearest Fe atom from O atom (Fe2), the occupancy is almost recovered to pure hcp-Fe.

The partial density of state (PDOS) of $Fe_9O$ also clearly shows the interaction between O atom and neighboring Fe atoms (Fig. 2C). The PDOS of Fe atoms in $Fe_xO$ are remained to be similar, except for that of Fe1 atom (red line). Spectral function obtained from DFT+DMFT calculation also exhibits equivalent behavior (Fig. S2). From Fe2 atom, PDOS is almost identical to DOS of hcp-Fe, which is consistent with occupancy analysis. At the Fermi level ($E_F$), Fe1 atom has a larger density than other Fe atoms. Due to the charge transfer, Fe1 bands are shifted upward, making flat bands near the $E_F$ (Fig. S2 and S3). The peak at around -11 eV observed in PDOS of both O and Fe1 atom indicates the strong hybridization between them. The crystal orbital Hamilton population (COHP) analysis between O and Fe1 atom can give insight on bonding properties (Fig. 2D) *(29)*. The overall shape of COHP indicates covalent bonding between O and Fe1 atom, rather than just ionic bonding between them. A notable bonding state is also shown in COHP at -11 eV.

Fig. 2E shows the relative lattice constant change with respect to oxygen contents. For the direct comparison, the $c$ lattice constant of each $Fe_xO$ ($x$=5~9) compound is renormalized by the number of atoms in the unit cell along $c$ direction. Due to the interaction between Fe and O atoms, $c$ lattice constant is affected sensitively by the oxygen contents. The $c$ lattice constant decreases ~ 5% at 10% oxygen content ($x$=9 case). $a$ lattice constant is less sensitive to the oxygen content and it increases as oxygen content increases Thus, the existence of O atoms in hcp-Fe enhances an anisotropic structural distortion not only along $c$-direction but in $ab$ plane. The bonding between Fe and O atom explains the elastic anisotropy at IC. Previous studies on the pure hcp-Fe show that the $c/a$ ratio becomes ideal as temperature increases *(3, 4)*, however, the anisotropy can be induced in the presence of a small amount of oxygen. Due to the strong interaction between Fe and O, $c/a$ ratio of hcp-Fe motif can be anisotropic even at high temperature. Although $Fe_xO$ exists as polycrystal at IC condition with a globally preferred orientation, it can induce the elastic anisotropy observed in the experiment. Therefore, the existence of oxygen is an important key to understanding the elastic anisotropy at IC conditions.

Pure hcp-Fe shows a dip feature in its spectral function at the $E_F$ (Fig. S2, 3). Due to the charge transfer and interaction between Fe1 and O atoms in $Fe_9O$, however, flat bands occur near the $E_F$. As pointed out in the previous studies *(7, 9, 30)*, a van Hove singularity enhances the correlation strength of systems, including bcc-Fe. Flat bands near the $E_F$ are well observed in $Fe_xO$, especially for $x \geq 7$ (Fig. S3). In addition, Fe1-Fe1 $3d$ orbitals overlap is strongly reduced because of oxygen, making the system more incoherent. Hence, we can expect the enhanced electron correlation effect in Fe$_x$O compared to pure hcp-Fe.

The resistivity from electron-electron scattering, $\rho_{e\text{-}e}$ of hcp-Fe and Fe$_9$O are computed with DFT+DMFT (Fig. 3A). To compare with the previously calculated result, we adopt the IC density for hcp-Fe and keep the density ratio between hcp-Fe and Fe$_9$O obtained from 0 K DFT calculation for Fe$_9$O. The calculated $\rho_{e\text{-}e}$ of hcp-Fe well agrees with the previous theoretical studies *(10)*. $\rho_{e\text{-}e}$ of Fe$_9$O is calculated to be higher than hcp-Fe and it monotonically increases with temperature, reaching almost doubled value around 5800 K than hcp-Fe. The resistivity increasing rate of Fe$_9$O shows slowdown above ~4600 K and that of hcp-Fe keeps increasing up to 7000K. We analyzed the temperature dependence of the inverse quasiparticle lifetime Γ (Fig. 3B). Γ/kT of Fe$_9$O is almost converged above ~4600K at which the resistivity increasing rate starts to decrease, signaling a fully incoherent regime *(9, 30)*. For hcp-Fe case, however, the fully incoherent regime does not appear up to 7000 K. Large $\rho_{e\text{-}e}$ (or Γ) and the reduced temperature scale of Fe$_9$O indicate that a small amount of oxygen enhances the electron correlation strength of hcp-Fe motif.

The total resistivity depending on the oxygen contents is shown in Fig. 3C. Diamond simples with blue color indicate computed $\rho_{e\text{-}e}$ using DFT+DMFT based on the structure obtained from DFT calculation at 300 GPa. The resistivity from electron-phonon scattering, $\rho_{e\text{-}ph}$ is taken from a previous study exhibited by red squares with dashed line *(31)*. One can notice that $\rho_{e\text{-}e}$ and $\rho_{e\text{-}ph}$ are both enhanced by the oxygen content and $\rho_{e\text{-}e}$ part is more sensitively changed. Black circles with dashed line show the sum of $\rho_{e\text{-}e}$ and $\rho_{e\text{-}ph}$. The resistivity of Fe$_9$O is ~20% larger than that of hcp-Fe. Considering the previously suggested oxygen portion in the Earth's core (3~6 wt%) *(16)*, the total resistivity can increase about 20~40% compared to the pure hcp-Fe.

Oxygen effect on thermal conductivity can be estimated by assuming the linear relation between electrical conductivity, $\sigma(\rho^{-1})$ and thermal conductivity, $\kappa$ (the Wiedemann-Franz law). 20~40% enhancement of resistivity leads to 17~29% reduction of thermal conductivity. Previous theoretical estimates on the thermal conductivity of pure hcp-Fe at Earth's core condition is $\kappa$ = 150~200 W m$^{-1}$ K$^{-1}$ *(8, 32)*, which is too large to explain the thermal convection in the geodynamo. Including the oxygen effect, it can be reduced to 100~140 W m$^{-1}$ K$^{-1}$ which is consistent with a thermally convection driven dynamo *(33)*.

Finally, we investigated the elastic properties of hcp-Fe and Fe$_9$O under inner core conditions. Standard DFT calculation was first adopted to verify the oxygen effect clearly. The calculated density and velocity of hcp-Fe at 300 GPa agree with the previous theoretical results. For Fe$_9$O, which has 3.09 wt% oxygen, the density decreases by ~3.2 % while the compression velocity (V$_P$) increases by ~1.9 % (Table S1). The electron correlation effect on the elastic properties was also verified by using DFT+DMFT calculation. The bulk modulus of hcp-Fe at 300 GPa decreases ~3.4 % which results in ~1.8 % reduction of seismic velocity with the assumption that the shear modulus is also reduced by the same amount. However, the thermal lattice vibration effect is dominant over the electronic correlation effect. A previous ab-initio molecular dynamics (AIMD) showed that the compression velocity decreases ~11.3 % at 5500 K compared to the 0 K result (Table S1).

The elastic constant of hcp-Fe and Fe$_9$O obtained by using AIMD calculations are shown in Table S2. Our calculated elastic constants of pure hcp-Fe at 360 GPa and 6000 K has a good match with previous theoretical results *(34, 35)*. The compression (V$_P$) and shear (V$_S$) velocities in hcp-Fe and Fe$_9$O are calculated with the elastic constants and compared with the PREM data (Fig. 4). As is well known, the density of pure hcp-Fe is higher than the PREM data. Fe$_9$O (3.09 wt% oxygen) gives much better result for density indicating the possible existence of oxygen in Earth's inner core. At 360 Gpa and 6000 K, the density of $Fe_9O$ decreases by ~3.8 % while the $V_p$ increases by ~1.3 % compared to those of hcp-Fe, like the simple DFT results. Considering the small density difference between the PREM data and $Fe_9O$ (330~360 GPa), there can be slightly more oxygen than 3.09 wt% which is consistent with the previous estimation (3~6 wt% oxygen) based on the seismological model *(16)*.

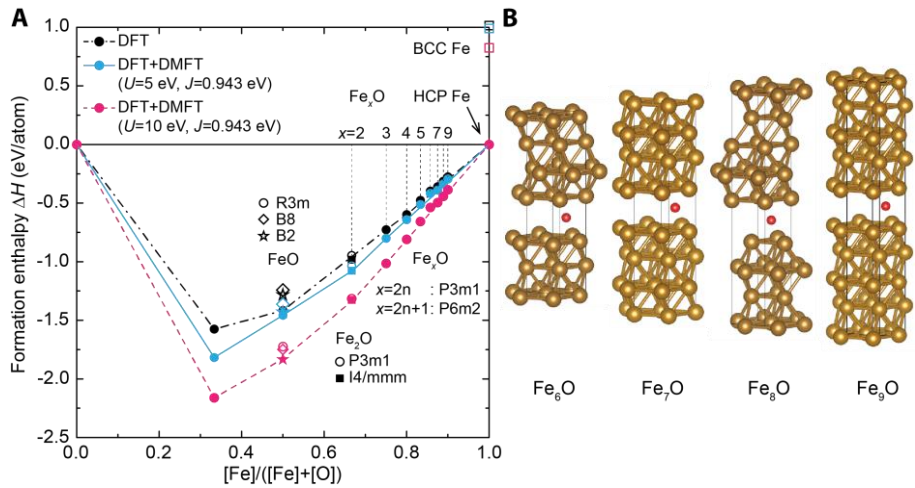

**Fig. 1. Crystal structure prediction of Fe$_x$O at the IC condtions.** (A) Convex hull plot for Fe-O system with the horizontal decomposition line into O and Fe. To verify the electron correlation effect on the formation energy, DFT+DMFT calculation was also employed. (B) Universal hexagonal crystal structure of Fe$_x$O ($x \geq 3$). Fe$_x$O can be understood as oxygen intercalated hcp-Fe. The small amount of oxygen can stabilize the hexagonal structure of Fe.

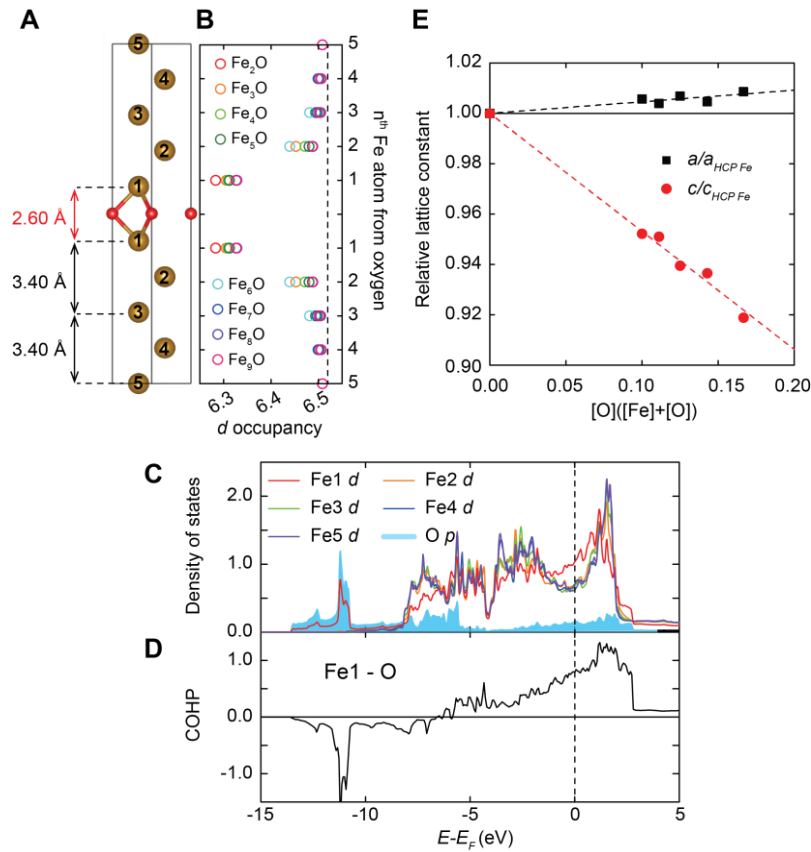

**Fig. 2. Bonding properties between Fe and O atoms.** (A) Crystal structure of $Fe_9O$ at 300 GPa. The bond length between Fe atoms above and below oxygen atom (Fe1-Fe1; red arrow) is shorter than the bond length between the second nearest Fe atoms (eg., Fe1-Fe3 or Fe3-Fe5; black arrows) (B) $d$ electron occupancy of Fe obtained from DFT+DMFT calculation. Due to the charge transfer from Fe1 to oxygen atom, $d$ occupancy of Fe1 atom is smaller than those of other Fe atoms (C) Partial density of states (PDOS) of $Fe_9O$. PDOSs of Fe2-Fe5 show a dip feature at the $E_F$ like pure hcp-Fe, while PDOS of Fe1 shows larger DOS at the $E_F$. (D) Crystal orbital Hamilton population (COHP) analysis between O and Fe1 atom. (E) Relative lattice constant depending on oxygen contents. The $c$ axis becomes shorter as oxygen contents increases.

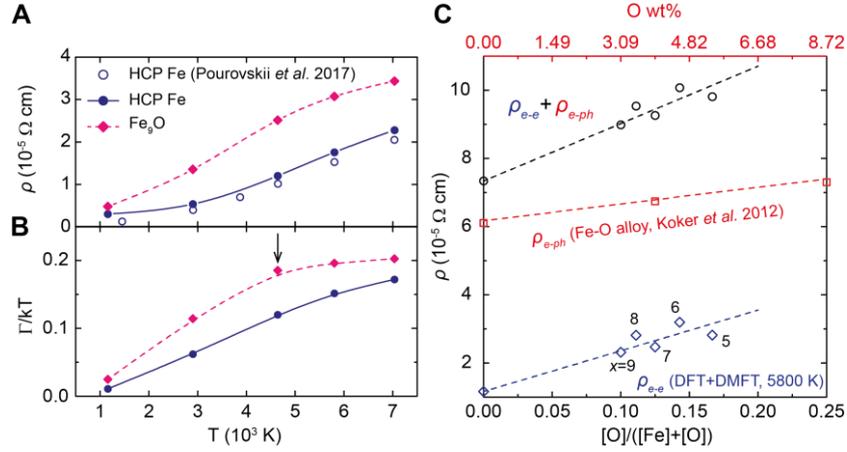

**Fig. 3. Resistivity of Fe$_x$O.** (A) Resistivity from *e-e* scattering part ($\rho_{e\text{-}e}$) from DFT+DMFT calculations. Our calculated result of hcp-Fe well agrees with the previous results. $\rho_{e\text{-}e}$ of Fe$_9$O is much higher than that of pure hcp-Fe. (B) Inverse quasiparticle lifetime Γ of hcp-Fe and Fe$_9$O. Γ/kT of hcp-Fe shows almost T linear behavior up to 7000 K, while that of Fe$_9$O is almost saturated at ~4600 K. (C) Total resistivity ($\rho_{e\text{-}e}$ +$\rho_{e\text{-}ph}$) depending on oxygen contents. $\rho_{e\text{-}ph}$ is taken from the previous studies. Although $\rho_{e\text{-}e}$ and $\rho_{e\text{-}ph}$ both increase as oxygen contents increase, $\rho_{e\text{-}e}$ is more sensitively affected by oxygen contents.

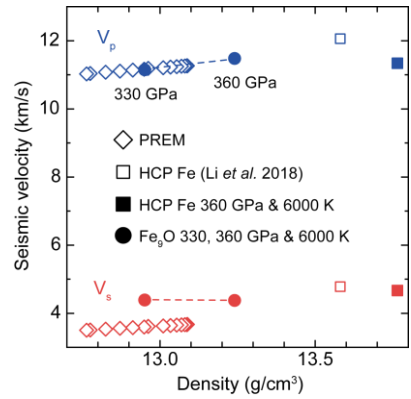

**Fig. 4. Seismic velocity of hcp-Fe and Fe$_9$O.** Diamond, square, and circle points indicate the seismic velocity of PREM, hcp-Fe, and Fe$_9$O, respectively. The density and velocity of pure hcp-Fe at 360 GPa is too high compared to the PREM data. For Fe$_9$O case, a small amount of oxygen suppresses the density and velocity, resulting in better agreement with RREM data.